\documentclass[aps,prd,amsmath,amssymb,superscriptaddress,altaffillsymbol, preprintnumbers,nofootinbib,twocolumn]{revtex4}
\pdfoutput=1
\usepackage{amsthm}
\usepackage{graphicx}
\usepackage{subfigure}
\usepackage{tikz}
\usepackage{color}
\newcommand{\beq}{\begin{eqnarray}}
\newcommand{\eeq}{\end{eqnarray}}

\newcommand{\bmp}{\noindent\begin{minipage}{16cm}}
\newcommand{\emp}{\end{minipage}\vskip 7mm} 

\usepackage{dcolumn}
\usepackage{bm}
\usepackage{bbm}
\usepackage{subfigure}
\usepackage{slashed} 

\usepackage{lipsum}

\theoremstyle{definition}

\theoremstyle{plain}

\usepackage{epsfig}

\usepackage{hyperref}
\definecolor{rossoCP3}{cmyk}{0,.88,.77,.40}
\definecolor{verdeCP3}{rgb}{0.09765625, 0.57421875, 0.1015625}
\definecolor{bluCP3}{rgb}{0, 0.23, 0.67}
\hypersetup{colorlinks, bookmarksopen, bookmarksnumbered,citecolor=bluCP3, linkcolor=bluCP3, pdfstartview=FitH, urlcolor=bluCP3}
\def\lsim{\mathrel{\rlap{\lower4pt\hbox{\hskip1pt$\sim$}}
    \raise1pt\hbox{$<$}}}                
\def\gsim{\mathrel{\rlap{\lower4pt\hbox{\hskip1pt$\sim$}}
    \raise1pt\hbox{$>$}}}                

\newcommand{\bea}{\begin{eqnarray}}
\newcommand{\eea}{\end{eqnarray}}

\newcommand{\ba}{\begin{eqnarray}}
\newcommand{\ea}{\end{eqnarray}}

\newcommand{\be}{\begin{eqnarray}}
\newcommand{\ee}{\end{eqnarray}}

\begin{document}
\title{Probing Light Dark Matter via Evaporation from the Sun}
%
\author{Chris Kouvaris}
\email{kouvaris@cp3-origins.net} 
\affiliation{{\color{black} CP$^{3}$-Origins} \& Danish Institute for Advanced Study {\color{black} DIAS}, University of Southern Denmark, Campusvej 55, DK-5230 Odense M, Denmark}


\begin{abstract}
Dark matter particles can be captured by the sun with rates that depend on the dark matter mass and the DM-nucleon cross section. However, for masses below $\sim 3.3$ GeV, the captured dark matter particles evaporate, leading to an equilibrium where the rate of captured particles is equal to the rate of evaporating ones. Unlike dark matter particles from the halo, the evaporating dark matter particles  have velocities that are not limited to values below the escape velocity of the galaxy. Despite the fact that high velocities are exponentially suppressed, I demonstrate here that current underground detectors have the possibility to probe/constrain low dark matter parameter space by (not)-observing the high energy tail of the evaporating dark matter particles from the sun. I also show that the functional form of the differential rate of counts with respect to the recoil energy in earth based detectors can identify precisely the mass and the cross section of the dark matter particle in this case.

\preprint{CP3-Origins-2015-022 DNRF90, DIAS-2015-22.}
 \end{abstract}

\maketitle

Despite the overwhelming amount of evidence from galaxy structure formation and cosmology in  favor of the existence of dark matter (DM), so far experiments have failed to conclusively detect directly or indirectly DM. A lot of scientific effort and resources  have been allocated in order to detect DM. In particular, underground detectors have imposed strict limits on the type of DM particles that can be viable, constraining the parameter space of  the mass and the cross section  
of DM interacting with baryons. The basic principle of direct detection is simple. A DM particle interacts with a nucleus of the detector deposing an amount of energy that is detectable. Different experiments apply different techniques on how they observe the recoil. However, direct detection rates have limitations. Obviously a small DM-nucleus cross section reduces the probability of interaction. Similarly, heavy DM particles have low number densities and consequently low flux. Low DM masses are also difficult to probe simply because the DM particle does not have enough energy to trigger the detector. This is true regardless the exposure that an experiment can achieve. No matter what is the velocity distribution of DM particles in the galaxy, their velocities are below the escape velocity of the galaxy. Therefore below a sufficiently small DM mass and a given detector energy threshold, no DM particle can be detected. This is the reason why low DM masses are not probed by direct DM searches. 

However, as I will demonstrate in this paper, it is possible to probe lighter DM masses with current detectors and energy thresholds due to a flux of DM particles that after being captured by the sun,  leak out via evaporation. These particles can arrive in earth with energies that are high enough to produce a detectable recoil. Therefore not only it is possible to probe relatively lighter DM, but additionally the spectrum has features that distinguish it clearly from heavier DM candidates. The possibility of detecting DM particles that have been captured by the sun, on the earth, has been studied in the past~\cite{Damour:1998rh,Damour:1998vg}. In these two seminal papers, the flux of  DM particles bound to the sun and having orbits that can reach the earth was estimated. Damour and Krauss considered particles that have been captured by the outer layers of the sun and due to perturbations from other planets, the orbits evolved to eliptical ones that do  not cross the sun anymore and therefore do not lose further energy. Particles in these orbits can accumulate for billions of years and since some of the orbits can reach the earth, these particles are potentially detectable. Additionally, this scenario has been studied numerically~\cite{Peter:2009mi,Peter:2009mm}. It should be emphasized here that this present paper studies a fundamentally different scenario. Instead of looking at loosely bound DM particles that have orbits that cross the earth, I focus on light particles that have had the time to thermalize with nuclear matter inside the sun. The tail of the distribution of these particles corresponds to velocities above the escape velocity of the sun and this is exactly the spectrum of particles that I consider here.

Generally, the number of DM particles in the sun $N$ is determined by 
\be
\frac{dN}{dt}=F-C_eN-C_a N^2, \label{rate}
\ee
where $F$ is the capture rate, and $C_{e,a}$ are coefficients related to the evaporation and annihilation of DM respectively. From the above equation it is clear that if $C_e^2>>C_a F$, and $C_e^{-1}$ is much smaller than the age of the solar system, evaporation dominates the whole process and an effective equilibrium between the accretion rate $F$ and evaporation has been established by now. In other words, if the above conditions are satisfied, DM particles leak out from the sun with the same rate as they are captured. Let us first estimate the capture rate~\cite{Press:1985ug,Gould:1987ir,Kouvaris:2010jy,Catena:2015uha}.  The capture rate is
\be
F=\frac{8 \pi^2}{3}\frac{\rho_{dm}}{m_{\chi}} \left (\frac{3}{2 \pi v_0^2} \right )^{3/2} G M_{\odot} R_{\odot} v_0^2 \sum_i(1-e^{-3E_i/v_0^2})f_i, \label{capture}
\ee
where $M_{\odot}$ and $R_{\odot}$ are the mass and the radius of the sun, $v_0$ the velocity dispersion of DM in our galaxy, $\rho_{dm}$ and $m_{\chi}$ the local DM density and the DM mass respectively. The sum runs over all different chemical elements present in the sun. I am going to consider for simplicity only hydrogen and helium. $E_i$ is the maximum energy per DM mass that can lead to a capture due to a collision of DM with element $i$ and is given by $E_i=\gamma_i G M_{\odot} /R_{\odot}(1-\gamma_i)$ where $\gamma_i=2m_{\chi}m_{Ni}/(m_{\chi}+m_{Ni})^2$ is the average fraction of energy that the DM particle loses after colliding with a nucleus $N_i$. Note that for energy larger than $E_i$, even if the particle scatters inside the sun will lose on average an amount of energy that is not sufficient to bind the particle gravitationally to the sun. Finally $f_i$ represents the probability that scattering will take place. $f_p=0.89 \epsilon_p \sigma_p/\sigma_{crit}$, if  $\epsilon_p \sigma_p/\sigma_{crit}<1$ and 1 if  $\epsilon_p \sigma_p/\sigma_{crit}>1$, where $\epsilon_p$ is the mass fraction of hydrogen in the sun that is taken to be 0.75. For helium, $f_{He}=0.89 \times 4\epsilon_{He} \sigma_p/\sigma_{crit}$, if  $4\epsilon_{He} \sigma_p/\sigma_{crit}<1$ and 1 if  $4\epsilon_{He} \sigma_p/\sigma_{crit}>1$, where $\epsilon_{He}=0.24$. $\sigma_{crit}=m_pR_{\odot}^2/M_{\odot}\simeq 4 \times 10^{-36}\text{cm}^2$ is roughly speaking the cross section above which every particle that will cross the sun will scatter.
Note that everything is expressed in terms of the DM-proton cross section $\sigma_p$. I consider spin-independent interactions and therefore the DM-helium cross section will be  $\sigma_{He}=\sigma_p (\mu_{He}^2/\mu_p^2) A^2$, where $\mu_i$ corresponds to the reduced mass of DM with nucleus $i$ and $A=4$ for helium. Since I am interested in low DM mass, $\sigma_{He}\simeq 16 \sigma_p$.

Now let us focus on evaporation. This effect has been studied extensively in the case of the sun~\cite{Spergel:1984re,Faulkner:1985rm,Krauss:1985aaa,Griest:1986yu,Gould:1987ju}. Unless one assumes unreasonably high DM annihilation cross section, it has been shown~\cite{Griest:1986yu,Gould:1987ju} that for DM masses below $\sim 3.3$ GeV, DM particles get effectively evaporated out of the sun. In this case the steady state solution of Eq.~(\ref{rate}) will give an equilibrium between captured and evaporated DM particles. Although the exact formula of $C_e$ has been estimated~\cite{Griest:1986yu,Gould:1987ju}, it will not be needed  here. As long as I consider particles below 3.3 GeV, 
the rate of evaporation will be equal to that of capture. Therefore the overall evaporation rate will be given by Eq.~(\ref{capture}). Let us now determine the spectrum of the evaporating DM particles. In general there are two possibilities. If the DM-nucleon cross section is large and the mean free path of the DM particle small, the captured population of DM will thermalize fast with nuclei and the DM distribution will be a Maxwell-Boltzmann one with a DM temperature equal to the one of the star at a particular position. However, if the DM-nucleon cross section is small, captured DM interacts over several orbits and therefore there is no single temperature that picks up. The distribution is not a Maxwell-Boltzmann one, but it can be approximated as one although the DM ``effective" temperature is different from that of the star~\cite{Spergel:1984re,Gould:1987ju}. This approximate distribution should look like
$f(v,r) \sim \exp[-E(v,r)/kT_{\chi}]$,
where $T_{\chi}$ is the ``effective" temperature of the DM, and $E(v,r)=m_{\chi}v^2/2+m_{\chi}V(r)$ is the total energy, $V(r)$ being the gravitational potential as a function of $r$ inside the sun. I would like to estimate the spectrum of evaporating DM particles now. From this point of view, the details of the spatial dependence of the  distribution are irrelevant since I consider particles that are at distance $R_{\odot}$ from the center of the sun with a velocity higher than the escape velocity of the sun. Therefore the spectrum of evaporating DM particles will be given by
\be 
f(v)=A e^{-\frac{m_{\chi} v^2}{2k T_{\chi}}},~~v>v_s, \label{boundary}
\ee
where $A$ is a constant to be determined and $v_s=(2G M_{\odot}/R_{\odot})^{1/2}$ is the escape velocity from the surface of the sun. However, this spectrum of evaporating DM particles does not remain the same when the DM particles arrive on earth. Let us find the spectrum of velocities at the earth $f(r,v)$ by using the collisionless Boltzmann equation
\be
\frac{\partial f}{\partial t}+v_r\frac{\partial f}{\partial r}+\frac{F_r}{m_{\chi}}\frac{\partial f}{\partial v_r}=0. \label{boltzmann}
\ee
Due to the isotropy of the problem, $\partial f/\partial \theta=\partial f/\partial \phi=0$.  I am interested in a steady state solution and therefore $\partial f/\partial t=0$. $F_r/m_{\chi}=-GM_{\odot}/r^2+GM_{\oplus}/(\ell -r)^2$ is the force due to gravity from the sun and the earth, where $r$ is the distance from the center of the sun, $M_{\oplus}$ is the mass of the earth, and $\ell$ is the distance between the sun and the earth. Note that $F_{\theta}=F_{\phi}=0$. The generic solution of Eq.~(\ref{boltzmann}) is $f(v_r^2-2GM_{\odot}/r-2GM_{\oplus}/(\ell-r))$. This solution should match the boundary distribution at the surface of the sun $f(R_{\odot},v)$ given by Eq.~(\ref{boundary}). Upon using this boundary condition, the distribution in earth is
\be
f(\ell,v)=A e^{-\frac{m_{\chi}}{2kT_{\chi}}(v^2+v_s^2-v_e^2)},~~v>v_e, \label{earth_distri}
\ee
 where $v_e=(2GM_{\odot}/\ell +2G M_{\oplus}/R_{\oplus})^{1/2}$ and $R_{\oplus}$ is the radius of the earth. I have omitted negligible terms of the order $\mathcal{O}(GM_{\oplus}/\ell)$. Note here that in Eq.~(\ref{earth_distri}), $v^2=v_r^2+v_{\theta}^2+v_{\phi}^2$, where the values of $v_{\theta}$ and $v_{\phi}$ are equal to the ones at the boundary of $r=R_{\odot}$ of Eq.~(\ref{boundary}), since Eq.~(\ref{boltzmann}) does not involve derivatives of them. DM particles that evaporated from the sun arrive in the earth almost radially. This is because the ratio $v_{\theta}/v_r$ of a particle arriving in the earth varies from 0 to a maximum value of $R_{\odot}/\ell\simeq 0.0046$. The total flux of evaporating DM particles arriving in the earth is
\be 
\int_{v_e}^{\infty}f(\ell,v) v d^3v=\frac{F}{4 \pi \ell^2}. \label{cond}
\ee
Recall that $d^3v=v^2dvd\cos\theta d\phi$ and as  mentioned above the solid angle integral part does not extend to the full $4\pi$ but it is constrained to the value mentioned above. The product of $A$ with the angular integration part is determined by Eq.~(\ref{cond}), thus leading to the following flux of evaporating DM on earth
\be
\mathcal{F}_l=C e^{-\frac{m_{\chi}}{2kT_{\chi}}(v^2+v_s^2-v_e^2)}v^3,~~v>v_e,
\ee
where the constant $C$ is 
\be
C=\frac{F}{4 \pi \ell^2}\left (\int_{v_e}^{\infty} e^{-\frac{m_{\chi}}{2kT_{\chi}}(v^2+v_s^2-v_e^2)}v^3 dv\right )^{-1}.
\ee

Let us now estimate the number of counts registered in an underground detector taking into account both the flux of evaporating DM and regular DM halo particles. The differential rate of counts per recoil energy is
\begin{widetext}
\be
\frac{dR}{dE_R}=N_T\left [\int_{v_{min}}^{\infty}\frac{d\sigma}{dE_R}\mathcal{F}_l(v) dv+\frac{\rho_{\chi}}{m_{\chi}}\int_{v_{min}}^{v_{esc}}\frac{d\sigma}{dE_R}f(v)vd^3v \right ], \label{ratefinal}
\ee
\end{widetext}
where $N_T$ is the number of targets in the detector, $\rho_{\chi}=0.3\text{GeV}/\text{cm}^3$ is the local DM density, $v_{esc}=550\text{km}/\text{sec}$ is the escape velocity of our galaxy and $v_{min}=\sqrt{m_N E_R/2\mu^2}$ is the minimum velocity required to produce a recoil $E_R$ in a DM collision with a nucleus of mass $m_N$ ($\mu$ being the reduced mass between DM and nucleus). For the distribution $f(v)$, a truncated Maxwell-Boltzmann function up to $v_{esc}$ of the form $f(v)=\mathcal{N}\exp[-(\vec{v}+\vec{v}_b)^2/v_0^2]$ is used, where $\mathcal{N}$ is a normalization constant~\cite{Savage:2006qr,Kouvaris:2015xga}. $v_b$ is the velocity of the earth with respect to the rest frame of the halo. The value used here is $v_b=(232+0.489\cdot 30)~\text{km}/\text{sec}$, which is the velocity of the solar system plus  the rotational velocity of the earth around the sun (when the latter aligns maximally with the former). This value represents the best possible scenario for detecting halo DM particles. Note that the first term in the right hand side of Eq.(\ref{ratefinal}) corresponds to the evaporating DM particles that arrive in earth from the sun, while the second one is the usual rate from incoming DM halo particles. A crucial observation here is that although DM halo particles have an upper velocity of $v_{esc}$, the evaporating ones can have any energy by paying a price in an exponential suppression in the density. However this fact  has important consequences because for a given threshold in DM detectors, and since $v<v_{esc}$, there is a mass below which DM particles from the halo can never have energies that can trigger the detector no matter how large the exposure is. On the contrary, for DM particles that have been captured first by the sun and later evaporated, it is probable to detect the tail of their distribution since there is no upper velocity, if enough exposure is achieved.

So far the value of the ``effective" DM temperature $T_{\chi}$ has not been specified. This has been estimated in  e.g.~\cite{Spergel:1984re,Gould:1987ju} and it depends on the DM mass. However both papers gave results for DM masses above $\sim2$ GeV. In order to find $T_{\chi}$ in much smaller masses of interest, I implement the method presented in~\cite{Spergel:1984re}. Although as  mentioned earlier, the actual distribution of captured DM is not an exact Maxwell-Boltzmann, it can be approximated by such with a temperature $T_{\chi}$. $T_{\chi}$ can be estimated by demanding no net flow of energy from the nuclei of the sun to the DM particles once a steady state has been achieved. The condition can be written~\cite{Spergel:1984re} as
\begin{widetext}
\be
\int d^3r~n_p(r) \int d^3 v \exp \left (\frac{-E}{kT_{\chi}} \right ) \int d^3 v_p \exp \left (\frac{-m_pv_p^2}{2kT(r)} \right ) \sigma_p |\vec{v}_{\chi}-\vec{v}_p| \langle \Delta E \rangle=0, \label{main_temp}
\ee
\end{widetext}
where $n_p(r)$ and $T(r)$ are the number density of nuclei and the temperature of the star at radius $r$ respectively, $v_{\chi}$ and $v_p$ are the velocities of DM and nuclei, and $E$ is the total energy of DM (i.e. kinetic plus potential). $\sigma_p$ is the DM-proton cross section and $\langle \Delta E\rangle$ is the energy exchange in a DM-nucleon collision.
For simplicity I use a polytropic model of $n=3$ as an approximation for the sun. If $\phi(\xi)$ is the solution of the  $n=3$ Lane-Emden equation, $n_p(\xi)=n_p(0)\phi(\xi)^3$, $T(\xi)=T_c \phi(\xi)$ and $V(\xi)=(4kT_c/m_p)[1-\phi(\xi)]$, where $\xi$ represents a dimensionless radius defined as $\xi=\xi_1 (r/R_{\odot})$, $\xi_1=6.8968486$ being the first zero of $\phi(\xi)$. Eq.~(\ref{main_temp}) can be rewritten in terms of the dimensionless quantities $\tau=T_{\chi}/T_c$ ($T_c$ being the core temperature of the sun) and $\nu=m_{\chi}/m_p$ as
\begin{widetext}
\be
\int_0^{\xi_1} \phi(\xi)^3 \exp\left [\frac{4 \nu}{\tau}(\phi (\xi)-1) \right ] \left (\frac{\tau+\nu \phi (\xi)}{\nu} \right )^{1/2}[\tau-\phi (\xi)]\xi^2 d\xi=0. \label{tempe}
\ee
\end{widetext}
Apart from a minor typo, the above equation is the same as that derived in~\cite{Spergel:1984re}. For every considered DM mass,  I have solved numerically Eq.~(\ref{tempe}) in order to find the corresponding $T_{\chi}$.

\begin{figure}[h!]
\begin{center}
\includegraphics[width=.4
\textwidth, height=0.3 \textwidth
]{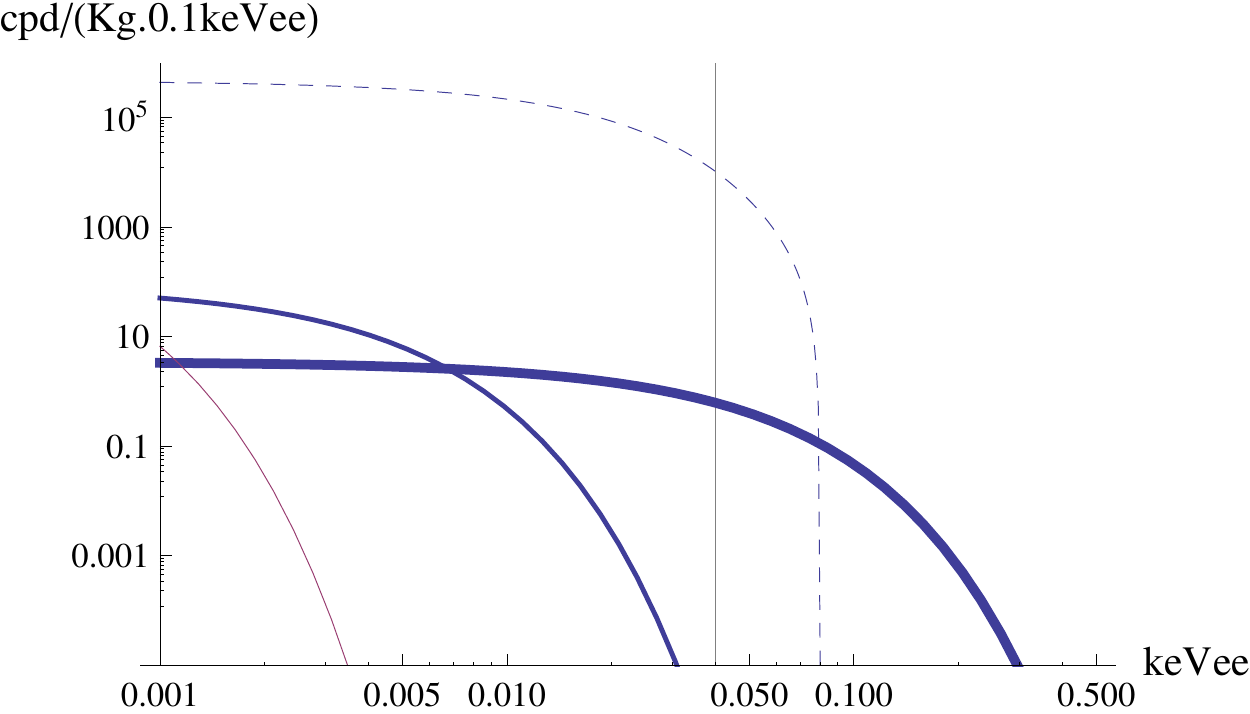}
\caption{Number of counts per 0.1 keVee recoil energy normalized to an exposure of 1 Kg$\cdot$day for a Si detector such the one used by DAMIC for DM-proton cross section of $\sigma_p=10^{-36}~\text{cm}^2$. A  flat efficiency of 0.17 for the detector at low masses has been used. The dashed line corresponds to halo DM counts of DM with a mass of 1 GeV. The solid lines correspond to counts coming from DM evaporating from the sun with masses 1 GeV (thick solid), 100 MeV (medium solid) and 10 MeV (thin solid). Note that for DM mass of 10 and 100 MeV, halo DM particles have not sufficient energy to recoil within the range shown in the plot. The vertical line represents the current threshold of DAMIC i.e. 40 eVee. }
\end{center}
\end{figure}

Fig.~1 shows some characteristic results that demonstrate why the spectrum of evaporating DM particles can enhance the chances for direct detection of light DM particles. It shows the rate of counts per recoil energy in bins of 0.1 keV, normalized to an exposure of 1 Kg$\cdot$day, for a Si detector like the one used in DAMIC. I quote DAMIC here because it is one of the experiments with the lowest recoil energy threshold. I have used a flat efficiency of 0.17 for the detector, deduced from~\cite{Barreto:2011zu}. Fig.~1 corresponds to DM-proton cross section of $\sigma_p=10^{-36}~\text{cm}^2$. For smaller cross sections, the rate of counts can be obtained by scaling the evaporating lines by $(\sigma_p/\sigma_{36})^2$ and the halo one by $\sigma_p/\sigma_{36}$, where $\sigma_{36}=10^{-36}~\text{cm}^2$. For the evaporating particles, one power of $\sigma_p$ comes from the capture rate in the sun and one from the detection  in the earth. I assume contact spin-independent  interactions (using a Helm form factor) in this paper, although the generalization for spin-dependent is trivial. In the plot three cases of DM masses are shown: 1 GeV, 100 MeV and 10 MeV. One can see that halo DM particles roughly below 1 GeV cannot be detected. The reason is that light DM particles do not have enough energy to create recoil energies above the threshold. Note that the number of counts for the halo DM in the plot, correspond to the best detection scenario, i.e. the period of the year where the earth rotational velocity around the sun aligns maximally with the sun velocity in the rest frame of the halo. On the contrary, light DM that has been captured by the sun and has evaporated after thermalization, have no upper bound (apart from the speed of light). Although high velocities are exponentially suppressed, depending on the detector exposure, such particles can be detected. Fig. 1 shows that slightly lower thresholds in direct detection can set limits down to DM masses of 10 MeV. This is impossible for halo DM particles that have no chance to be detected even with a  threshold of order of eV.

The spectrum of evaporating DM particles not only can probe/constrain parameter space that is not accessible by observing  halo DM particles, but it can potentially determine accurately the DM mass and DM-nucleon cross section. As it can be seen in Fig. 1 for the case of $m=1$ GeV, there is a change in the shape of the rate of counts in the detector. For low recoil energies, halo DM particles dominate the counts. However the counts produced by halo DM particles drop sharply at the maximum recoil energy $\gamma_{Si} m (v_{esc}+v_b)^2$ ($\gamma_{Si}$ refers to the nucleus of Si and it is defined below Eq.~2). Above this specific value of recoil energy only evaporating DM particles contribute to the counts. Therefore this drop in the number of counts per energy can lead to the exact determination of the DM mass and the DM-nucleon cross section.

Some comments are in order here. One should make sure that there is enough time for the DM particles to thermalize with the interior of the sun. The issue has been addressed in~\cite{Kouvaris:2010jy}  where it was shown that the characteristic time scales are of order of a year or smaller (for the bulk of the DM orbits). The second comment is related to DM annihilation. The spectrum of evaporating DM particles shown here is valid whether one considers asymmetric DM or thermally produced symmetric DM. As it was argued in~\cite{Gould:1987ju}, for a DM mass  of 3 GeV, the evaporation rate is equal to the annihilation one (with an annihilation cross section that of the weak interactions). For every 0.3 GeV below that mass value, the annihilation signal is suppressed by a factor of 100 compared to evaporation, practically eliminating the annihilation below $\sim$ 2 GeV.  Therefore the spectrum of the evaporating particles predicted here does not depend on the nature (asymmetric or symmetric) of DM. 

I should also mention that the evaporating DM particles can create an annually modulated signal with a different phase from the one of the halo DM particles. Here, the modulation is due to the small changes in the distance of the earth to the sun between summer and winter. The perihelion (shortest distance) takes place around January 3 and the aphelion (largest distance) around  July 4. The small fluctuation in the distance creates a fluctuation in the DM flux arriving on earth, thus the annual modulation. The largest signal should be expected  around January 3. Therefore there is a phase difference by almost a $\pi$ with respect to the annual modulation of the halo DM particles.

Finally, one can study the same effect from evaporating DM particles from the earth. Although the capture rate of the earth is on average smaller by at least eight orders of magnitude, this can be counterbalanced by the fact that the flux is inversely proportional to the distance, thus earth is favored by a factor of $(\ell/ R_{\oplus})^2$. Additionally, the mass below which evaporation dominates is not much different from the case of the sun. Although $\sigma_{crit}$ for the earth might be a bit smaller compared to the sun, as well as the thermalization time slightly larger, the flux of evaporating DM particles is not significantly lower than the one coming from the sun. Despite this fact, there is a fundamental difference. The spectrum of the evaporating DM from the earth is dominated by the velocities close to the earth's escape velocity $\sim 11~\text{km}/\text{sec}$, which is small to create significant recoil  for light DM. This is why I do not examine evaporating DM from the earth here.

The author is supported by the Danish National Research Foundation, Grant No. DNRF90.

\end{document}